\documentclass[prd,twocolumn,superscriptaddress,floatfix,amsmath,amssymb,amsfonts,longbibliography,nofootinbib]{revtex4-2}
\usepackage{amsmath,amssymb,color,graphicx,bm,hyperref,mathrsfs}
\usepackage{physics}
\usepackage{verbatim}

\newcommand{\E}[1]{Eq.~(\ref{#1})}

\definecolor{red}{rgb}{0.8,0,0}
\definecolor{RED}{rgb}{0.8,0,0}
\definecolor{violet}{rgb}{0.4,0,0.4}
\definecolor{green}{rgb}{0,0.5,0.0}
\definecolor{GREEN}{rgb}{0,0.5,0.0}
\definecolor{navy}{rgb}{0.0,0.0,0.6}
\definecolor{orange}{rgb}{0.8,0.2,0.0}
\definecolor{blue}{rgb}{0.3,0.0,0.8}

\begin{document}
\title{\textbf{Thermodynamic parameters of fluids on conformally connected spacetimes}}
\author{Bhera Ram}
\email{bhera.ram@iitg.ac.in}
\author{Bibhas Ranjan Majhi}
\email{bibhas.majhi@iitg.ac.in}
\affiliation{Department of Physics, Indian Institute of Technology Guwahati, Guwahati 781039, Assam, India.}

\begin{abstract}
Local {\it thermal equilibrium} generally implies the absence of heat flux within a fluid. We find the relations between a set of thermodynamic variables of a fluid on a general spacetime and those defined on a conformally connected spacetime, assuming both descriptions are at thermal equilibrium. The scaling relations appear to be consistent with Dicke's heuristic argument and the previous analysis done on the basis of various restrictions. Within the present framework, it is observed that the satisfaction of Klein's law on one of the spacetimes implies its validity on the other one. Moreover, our analysis bypasses some of the imposed restrictions and thereby reveals the generality of the earlier predictions.
These relations are further shown to preserve the geometric structure of thermodynamics, known as \textit{geometrothermodynamics}, such that the associated metrics on two conformally connected spacetimes are themselves conformally related. This provides an alternative consistency check as well as a distinct geometric interpretation of the relations.

\end{abstract}                                                                                               
\maketitle

\section{Introduction}
Extending the thermal physics of fluids from a non-relativistic to a relativistic domain and to a curved background is both interesting and difficult. As an example, it is well known that Zeroth's law gets modified in the presence of gravity. Tolman and Ehrenfest \cite{Tolman:1930zza,Tolman:1930ona} showed that for static spacetime, $T (x^\alpha)\sqrt{-g_{00}(x^\alpha)}$ remains constant at the thermal equilibrium. This is known as the Tolman-Ehrenfest (TE) relation. Here $T(x)$ is the locally measured temperature of the fluid with respect to static observers, and $g_{00}$ is the time-time component of the static metric. The same was obtained for stationary cases as well, using fluid flow along the timelike Killing vector \cite{PhysRev.76.427.2}. Other choices of fluid flows have been discussed by Santiago and Visser \cite{Santiago:2018lcy}. Later, Klein showed that $({\mu}/{T})$ remains constant for static spacetime known as Klein's law, where $\mu (x)$ is the local chemical potential \cite{RevModPhys.21.531}. Consequently, various approaches and extensions have been proposed in the literature  \cite{WJ,PhysRevD.12.956,Sorkin:1981wd,Gao:2011hh,PhysRevD.85.027503,Green:2013ica,Shi:2021dvd,Shi:2022jya,Roupas:2013fct,Roupas:2014sda,Roupas:2018abp,Roupas:2014hqa,Rovelli:2010mv,Lima:2019brf,Lima:2021ccv,RAM2025169963}. Furthermore, in absence of thermal equilibrium the TE relation has gradient due to heat flux \cite{Miranda:2024xda}, which has been more generally discussed in \cite{PRDRAM}. 

In this work, we want to address a completely different aspect of the thermodynamics of fluids on a general background, which are solutions of Einstein's gravitational theory (our restriction to Einstein's theory will be made clear later on). Particularly, we are interested in finding possible relations between the set of fluid thermodynamic variables defined on a spacetime and those defined on its conformally connected background (e.g. a pair of such solutions of Einstein's gravity has been suggested in \cite{Sultana:2005tp}).  
  Specifically, suppose a fluid is in thermal equilibrium in the metric $g_{ab}$, with an equilibrium temperature $T$ and chemical potential $\mu$. In that case, we would like to know whether the equilibrium temperature and chemical potential can be determined for a conformally connected background defined by $\Tilde{g}_{ab}(x)= \Omega^2(x) g_{ab}(x)$, provided the fluid on $\tilde{g}_{ab}$ is also in thermal equilibrium. This is a natural question one can ask, and it has been posed in literature. 
  
  Initially, Dicke gave some heuristic arguments regarding the scaling of physical quantities under conformal transformations \cite{Dicke}. Dicke showed that under conformal scaling, mass scales as $\tilde{m} = \Omega^{-1} m$, while time and length scale as $\tilde{\ell}, \tilde{t} \sim \Omega$. Since the speed of light remains unchanged and energy scales like mass, the temperature also follows $\tilde{T} \sim \Omega^{-1}$. This suggestion has been tested in physical systems, like for the thermodynamic parameters of black hole horizon, explicitly (e.g. see \cite{Saida:2007ru,Faraoni:2014lsa,Majhi:2014hpa,Majhi:2014lka,Majhi:2015tpa,Prain:2015tda,Bhattacharya:2016kbm,Cote:2019fkf}) and the results have appeared to be consistent with Dicke's argument. This particular information is very important and handful since one set of thermodynamic variables can be obtained by knowing the other set.

  In a recent study, Faraoni and Vanderwee \cite{Faraoni:2023gqg}, motivated by the physical considerations of the cosmic microwave background in FLRW universes and Dicke's arguments, addressed this issue and demonstrated that the temperatures of fluids in the conformally connected backgrounds vary inversely with the scale factor. Although this analysis unveils various information and provides an important step towards this direction, it suffers from crucial limitations.  
Their analysis is limited to the fact that the seed metric is static. Further, they have used several explicit thermodynamic expressions for ideal fluids. Moreover, it is based on Eckart's first-order formalism \cite{Eckart:1940te}, which is known to suffer from issues of acausality and instability. Thus, one wonders how much of these restrictions can be relaxed, extending the validity of these results.

The present work is in this particular direction. To avoid the issue of acausality and other issues of earlier first-order fluid descriptions, like Eckart \cite{Eckart:1940te} and Landau \cite{Landau}, here we consider the more viable first-order formalism, recently proposed by F. S. Bemfica, M. M. Disconzi and J. Noronha \cite{PhysRevX.12.021044} (popularly called as BDNK theory). This has been formalised through the works done in \cite{PhysRevD.98.104064,PhysRevD.100.104020} and  \cite{Kovtun:2019hdm,Hoult:2020eho}.
The constitutive relations in BDNK first-order formalism that give the baryon current and the energy-momentum tensor are 
\begin{eqnarray}
   &&J^a=n u^a~, 
   \label{eq:A1}
\\   
  &&T^{a b}=(\varepsilon+\mathcal{A}) u^a u^b+(p+\Pi) \Delta^{a b}-2 \eta \sigma^{a b}
  \nonumber
  \\
&&\hspace{1cm}+u^a {q}^b+u^b{q}^a~.
   \label{eq:A2}
\end{eqnarray}
where the expressions for $\mathcal{A}$, $\Pi$, $q^a$ and $\sigma^{a b}$ are as follows:
\begin{eqnarray}
&&\mathcal{A}=\tau_{\varepsilon}\left[u^a \nabla_a \varepsilon+(\varepsilon+p) \nabla_a u^a\right]~,
\label{eq:A3}
\\
&&\Pi=-\zeta \nabla_a u^a+\tau_p\left[u^a \nabla_a \varepsilon+(\varepsilon+p) \nabla_a u^a\right]~,
\label{eq:A4}
\\
&&{q}_b = \frac{  \sigma T(\varepsilon+p)}{n} \Delta_b ^a \nabla_a(\frac{\mu}{T})
\nonumber
\\
&&+\tau_q\left[(\varepsilon+p) u^a \nabla_a u_b+\Delta_b ^a \nabla_a p\right]~,
\label{eq:A5}
\\
&&{\sigma}^{a b}=\frac{1}{2}\Delta^{a c}\Delta^{b d}\left( \nabla_c u_d + \nabla_d u_c - \frac{2}{3}\Delta_{c d}\Delta^{e f}\nabla_e u_f \right)~,
\nonumber
\\
&&=\frac{1}{2}\left(\nabla^a u^b + \nabla^b u^a + u^a \dot{u}^b +  u^b \dot{u}^a  - \frac{2}{3}\Delta^{a b}\nabla_e u^e\right).
\label{eq:A6}
\end{eqnarray}
In the above expressions, $\varepsilon$, $s$, $p$, $n$, $T$, and $\mu$ are equilibrium thermodynamic variables -- energy density, entropy density, pressure, number density, temperature, and chemical potential, respectively, connected via the Euler relation 
\begin{equation}
\varepsilon + p = T s + \mu n~.
\label{ER}
\end{equation}
Further, $u^a$ is a normalized time-like vector (\textit{i.e.} $u_a u^a = -1$) called the flow or fluid velocity, and $\Delta_{a b} = g_{a b} + u_a u_b$ is a projector onto the space orthogonal to $u^a$. $\dot{u}^a$ denotes the acceleration of the fluid flow and is given by $\dot{u}^a=u^b\nabla_bu^a$.
Also, the out-of-equilibrium corrections to the energy-momentum tensor are given by the energy density correction $\mathcal{A}$, the bulk viscous pressure $\Pi$, and the heat flux $q^a$, respectively. Furthermore, $\sigma_{ab}$ denotes the traceless shear tensor, with $\zeta$, $\sigma$, and $\eta$ being the coefficients of bulk viscosity, heat conductivity, and shear viscosity. This description is consistent with the required properties \cite{PhysRevX.12.021044}. Particularly, the local version of the entropy increase theorem is well satisfied upon using the on-shell conditions given by
\begin{align}
\nabla _a S^a &= 2\eta\sigma_{a b}\sigma^{a b} + \zeta \frac{(\nabla_a u^a)^2}{T} \notag \\
&\quad + \sigma T \left[\Delta^{a b} \nabla_b\left(\frac{\mu}{T}\right)\right]^2
+ \order{\partial^3}~,
\label{entropy}    
\end{align}
which is non-negative when $\eta$,$\zeta$,$\sigma$ $\geq 0$, where $S^a = s u^a + \frac{q^a}{T} + \mathcal{A}\frac{u^a}{T}$ is the entropy current (for a detailed discussion, see \cite{RAM2025169963}). Now, since the stability and causality of the above theory have been tested only within Einstein's theory of gravity \cite{PhysRevX.12.021044}, the application of the formalism can be trusted to backgrounds which are solutions of this particular theory of gravity.

Since our calculation is based on BDNK theory, the results are consistent with the laws of physics and can, therefore, be trusted. Apart from this, there are a few important aspects of our analysis.
We find that several of the restrictions employed in earlier analysis can be relaxed, proving the generality and wider applicability of these outcomes. These are as follows.
We find that the temperature in the conformally connected background varies inversely with the scale factor as discussed in their analysis \cite{Faraoni:2023gqg}. However, it appears that such a relation is true for any two conformally connected backgrounds (these need not be static or stationary) which are solutions to Einstein's equations. Moreover, the fluids can be viscous in both spacetimes. Additionally, we find the relation for chemical potentials and also describe the relations between other equilibrium thermodynamic variables for the two conformally connected backgrounds. As a corollary of these findings it is being observed that if the seed metric is static - implying Klein's law holds such that $\mu/T$ remains constant throughout the spacetime - then $\Tilde{\mu}/\Tilde{T}$ similarly remains constant in the conformally connected background, even if the latter background is neither static nor stationary. 

Further, by employing the framework of Geometrothermodynamics (GTD), developed by Quevedo \cite{Quevedo_2007, quevedo2011fundamentalsgeometrothermodynamics}, which formulates the thermodynamic properties of a system using the language of differential geometry and Legendre-invariant metrics, we find the geometric description of equilibrium thermodynamics on the corresponding backgrounds. Specifically, we show that the Legendre invariant metrics associated with two conformally connected backgrounds, upon applying the relations between the thermodynamic parameters, are themselves conformally related. This could be an alternative verification of the consistency of the relations among the equilibrium thermodynamic parameters. Moreover, the causal structure encoded in these metrics offers a novel perspective for interpreting Klein's law within the GTD framework.

However, it must be noted that these results are valid when both fluids are in thermal equilibrium (no flow of heat flux) on the respective spacetime. Our analysis is done in $(3+1)$-spacetime dimensions. Also, the signature of the spacetime metric is taken to be $(-,+,+,+)$.

\section{Fluid parameters in thermal equilibrium}
 Consider a dissipative fluid in the background $g_{ab}(x)$, described by (\ref{eq:A1}) -- (\ref{eq:A6}). We are interested in investigating how the equilibrium fluid variables on $g_{ab}(x)$ are related to those on a conformally connected background $\Tilde{g}_{ab}(x)$ given by 
 \begin{equation}
\Tilde{g}_{ab}(x) = \Omega^2(x)g_{ab}(x)~.
\label{BR2}
\end{equation}
In this case, the fluid description on $\tilde{g}_{ab}(x)$ is of the form (\ref{eq:A1}) -- (\ref{eq:A6}) with untilde variables replaced by tilde variables.
 Within this stable and causal first-order formalism, we demand that the fluid is in {\it local thermal equilibrium} (absence of heat flux) in both the backgrounds $\textit{i.e.}\,\, q^a=0$ and $\Tilde{q}^a=0$, so that local equilibrium temperature $T(x)$ and $\Tilde{T}(x)$ can be defined in both the backgrounds. To look for a relation between these two temperatures, we will proceed in the following way. We denote the untilde quantities for $g_{ab}$ while the tilde ones are defined on $\tilde{g}_{ab}(x)$.

The use of Euler's relation (\ref{ER}) along with first law of thermodynamics $\mathrm{d}\varepsilon = T\mathrm{d}s + \mu \mathrm{d}n$, yields
\begin{equation}
\mathrm{d} p= s \mathrm{d}T + n \mathrm{d}\mu~.
\label{D1}
\end{equation}
The covariant form of the above equation, with further use of the Euler relation, gives
\begin{equation}
\nabla_a p = \left(\frac{\varepsilon+p}{T}\right)\nabla_a T +  n T\,\nabla_a\left(\frac{\mu}{T}\right)~.
\label{D2}
\end{equation}
Moreover, the use of Eq. (\ref{D2}) in Eq. (\ref{eq:A5}) provides
\begin{multline}
{q}^a = \left(\frac{\sigma T(\varepsilon+p)}{n} + \tau_q n T\right) \Delta^{ab} \nabla_b\left(\frac{\mu}{T}\right)\\
+\tau_q\left({\varepsilon+p}\right)\Delta^{ab}\left[\nabla_b \ln{T}+\dot{u}_b\right]~,
\label{D3}
\end{multline}
which can be further rewritten as
\begin{multline}
{q}^a = \alpha \Delta^{ab} \nabla_b\left(\frac{\mu}{T}\right)+\beta \Delta^{ab}\left(\nabla_b \ln{T}+\dot{u}_b\right)~.
\label{D4}
\end{multline}
In the above we define $\alpha = {\sigma T(\varepsilon+p)}/{n} + \tau_q n T$ and $\beta = \tau_q\left(\varepsilon+p\right)$.
Similarly, in the conformally connected background given by Eq. (\ref{BR2}), the expression for the heat current is given as
\begin{eqnarray}
&&\Tilde{q}^a =\Tilde{\alpha} \Tilde{\Delta}^{ab} \Tilde{\nabla}_b\left(\frac{\Tilde{\mu}}{\Tilde{T}}\right) +\Tilde{\beta}\Tilde{\Delta}^{ab} \left(\Tilde{\nabla}_b \ln{\Tilde{T}} + \dot{\Tilde{u}}_b\right)~,
\label{D5}
\end{eqnarray}
where $\Tilde{\alpha}$ and $\Tilde{\beta}$ are given by,  $\Tilde{\alpha} = {\Tilde{\sigma} \Tilde{T}(\Tilde{\varepsilon}+\Tilde{p})}/\Tilde{n} + \Tilde{\tau_q} \tilde{n} \Tilde{T}$ and $\Tilde{\beta} = \Tilde{\tau_q}\left(\Tilde{\varepsilon}+\Tilde{p}\right)$.
Now, since we have $u^au_a=-1$ and $\Tilde{u}^a\Tilde{u}_a=-1$, then one normalization condition will imply the other one, provided we have 
\begin{equation}
\Tilde{u}^a = \frac{u^a}{\Omega}~,
    \label{BR4}
\end{equation}
which further gives $\Tilde{u}_a = \Omega {u}_a$.
Furthermore, one can easily obtain the relation between the accelerations, which is given by
\begin{equation}
\dot{\Tilde{u}}_a = \dot{u}_a + \Delta_{ab}\nabla^b\ln{(\Omega)}~.
\label{BR5}
\end{equation}

\subsection{Temperature and chemical potential}
Now, requiring thermal equilibrium in the seed background also ensures thermal equilibrium in the conformally connected background, we must have
\begin{equation}
\Tilde{q}^a = \Omega^{z_1}q^a~,
\label{imp}
\end{equation}
where $z_1$ is some real number. The above choice confirms that if $q^a$ vanishes at each event on $g_{ab}(x)$, then $\tilde{q}^a$ also vanishes at the corresponding spacetime event on $\tilde{g}_{ab}(x)$. This is legitimate as the present conformal transformation between the metrics does not change the coordinates. Moreover for $\Omega=1$ both $q^a$ and $\tilde{q}^a$ must be same. Then it leads to the following relations:
\begin{equation}
\Tilde{T} = \frac{T}{\Omega}~;
\label{BR6}
\end{equation}
and
\begin{equation}
\Tilde{\mu} = \frac{\mu}{\Omega}~,
\label{BR20}
\end{equation}
between temperature and chemical potential in both backgrounds, which we have explicitly demonstrated in Appendix \ref{App1}. The present analysis shows that the above results do not depend on the explicit value of $z_1$. As of now, this can be any non-vanishing real arbitrary number.
The above choices are very natural as we are restricting to the situations where $q^a=0$ implies $\tilde{q}^a=0$. 
The relation between the temperatures has been derived in the \cite{Faraoni:2023gqg}, but in their analysis, the seed metric $g_{ab}(x)$ is static, and they have used Eckart's first-order formalism, which, as mentioned, suffers from being acausal and unstable. Here, we have avoided these restrictions.
Moreover, combining Eq. (\ref{BR6}) and Eq. (\ref{BR20}), we obtain an expression relating the ratio of chemical potential and temperature in the two conformally connected backgrounds, given by
\begin{equation}
\frac{\Tilde{\mu}}{\Tilde{T}} = \frac{\mu}{T}~.
\label{BR9}
\end{equation}
Therefore, we find that demanding thermal equilibrium in both backgrounds, which are conformally connected, the temperature and chemical potential in $\Tilde{g}_{ab}$ are inversely proportional to the conformal factor. 

Furthermore, Eq. (\ref{BR9}) provides an interesting implication. Consider a special case where the seed background $g_{ab}(x)$ is static or stationary. In that case, we have Klein's law, which states $\mu/T$ is constant over the entire spacetime \cite{RevModPhys.21.531,Green:2013ica}. However, because of Eq. (\ref{BR9}), we find that $\Tilde{\mu}/\Tilde{T}$ remains constant in the conformally connected background, even though $\Tilde{g}_{ab}(x)$ is not necessarily static or stationary. 

\subsection{Other thermodynamic variables}
Let us now discuss the relations for the other thermodynamic variables. Apart from the above-discussed relations between temperature and chemical potential, we can further extend this analysis to obtain the relations between other fluid variables. On using the conservation for baryon number for the conformally connected metric,
we have
\begin{eqnarray}
  \int\mathrm{d}^4x\sqrt{{-\Tilde{g}}}\,\Tilde{\nabla}_a\Tilde{J}^a = \int\mathrm{d}^4x\,{\partial}_a\left(\sqrt{-\Tilde{g}}\,\Tilde{J}^a\right) = 0~.
    \label{BR12}  
\end{eqnarray}
On using Eq. (\ref{BR2}) and $\sqrt{-\Tilde{g}}=\Omega^4 \sqrt{-g}$, we have
\begin{eqnarray}
\int\mathrm{d}^4x\sqrt{{-\tilde{g}}}\,\tilde{\nabla}_a\tilde{J}^a 
    &=& \int\mathrm{d}^4x\,\sqrt{-g}\left\{\frac{1}{\sqrt{-g}}{\partial}_a \left(\Omega^4\sqrt{-g}\tilde{J}^a\right)\right\} \nonumber \\
    &=& 0~.
\label{S1}
\end{eqnarray}
Here we demand that the conservation of baryon current in one background will imply the conservation of the same in a conformally connected background. Thus Eq. (\ref{S1}) will imply
\begin{equation}
    \int\mathrm{d}^4x\sqrt{{-\Tilde{g}}}\,\Tilde{\nabla}_a\Tilde{J}^a = \int\mathrm{d}^4x\sqrt{{-g}}\,{\nabla}_a{J}^a = 0~,
    \label{S2}
\end{equation}
which yields
\begin{equation}
\Tilde{J}^a = \frac{J^a}{\Omega^4}~.
    \label{S3}
\end{equation}
Using Eq. (\ref{eq:A1}) for \( \tilde{g}_{ab} \), we find \( \tilde{J}^a = \tilde{n} \tilde{u}^a \). Substituting Eq. (\ref{S3}) and \( \tilde{u}^a = u^a / \Omega \), we obtain \( J^a = \Omega^3 \tilde{n} u^a \), which on comparison with Eq. (\ref{eq:A1}) for the seed metric yields
\begin{equation}
\Tilde{n}= \frac{n}{\Omega^3}~.
\label{BR15}
\end{equation}
Further, if we use the Euler relation for $\Tilde{g}_{ab}$, we have
\begin{equation}
\Tilde{\varepsilon}+\Tilde{p} = \Tilde{T}\Tilde{s}+ \Tilde{\mu}\Tilde{n}~,
\label{BR16}
\end{equation}
which on using Eq. (\ref{BR6}), Eq. (\ref{BR20}) and Eq. (\ref{BR15}), yields
$$\Tilde{\varepsilon}+\Tilde{p} = \frac{T\Tilde{s}}{\Omega}+ \frac{\mu\,{n}}{\Omega^4}. $$
Demanding that each term in Eq. (\ref{BR16}) must scale as $1/\Omega^4$ to yield the Euler relation (\ref{ER}) on $g_{ab}$, gives
\begin{eqnarray}
&&\Tilde{\varepsilon} = \frac{\varepsilon}{\Omega^4},\,\,\,\,\Tilde{p} = \frac{p}{\Omega^4},\,\,\,\,\Tilde{s} = \frac{s}{\Omega^3}.
\label{BR17}
\end{eqnarray}
The same relations are discussed for conformal transformations by Faraoni but for perfect fluids, \cite{Faraoni:2004Book}.

Also, as has been discussed at the end of Appendix \ref{App1}, one has
\begin{equation}
\frac{\Tilde{\alpha}}{\Tilde{\beta}} = \frac{\alpha}{\beta}~.
\label{E18}
\end{equation}
Using the expressions for the above quantities, we have
\begin{equation}
\frac{{\Tilde{\sigma} \Tilde{T}(\Tilde{\varepsilon}+\Tilde{p})}/\Tilde{n} + \Tilde{\tau_q} \tilde{n} \Tilde{T}}{\Tilde{\tau_q}(\Tilde{\varepsilon}+\Tilde{p})} = \frac{{\sigma T(\varepsilon+p)}/{n} + \tau_q n T}{\tau_q\left(\varepsilon+p\right)}~.
\label{Ex1}
\end{equation}
Now use of Eq. (\ref{BR6}), Eq. (\ref{BR15}) and Eq. (\ref{BR17}) into Eq. (\ref{Ex1}), we get
\begin{equation}
\left(\frac{\Tilde{\sigma}}{\Tilde{\tau}_q}\right) = \Omega^{-2}\left(\frac{\sigma}{\tau_q}\right)~.
\label{Ex2}
\end{equation}
This completes our analysis. 
Additionally, for more elaboration, we analyze a case where the fluid is in complete equilibrium in the background $g_{ab}$, and examine whether this equilibrium condition persists in the conformally connected background $\tilde{g}_{ab}$.

\subsection{Conservation of local entropy current in $g_{ab}$}
The increase in the entropy for the BDNK first-order formalism is given by Eq. (\ref{entropy}).
Now, if $\nabla_a S^a = 0$, implying that the fluid is in complete thermodynamic equilibrium—i.e., $\sigma_{ab} = 0$, $\Theta = \nabla_a u^a = 0$, and $q^a = 0$ in the seed background $g_{ab}$. Then, under those conditions, it can be shown that there exists a timelike Killing vector field for $g_{ab}$, given by $\xi^a = u^a / T$ \cite{RAM2025169963}. However, $\xi^a = \tilde{u}^a/\tilde{T}$ (obtained by using Eq. (\ref{BR4}) and Eq. (\ref{BR6})) is not a Killing vector for $\tilde{g}_{ab}$. 
Moreover, the relationships between the shear tensor and the expansion factor in both backgrounds are given by the following expressions:
\begin{equation}
\Tilde{\sigma}_{ab} = \Omega\,\sigma_{ab}~;
\label{B1}
\end{equation}
and
\begin{equation}
\Tilde{\Theta}= \frac{\Theta}{\Omega}+\frac{3\dot{\Omega}}{\Omega^2}~,
\label{B2}
\end{equation} 
where $\dot{\Omega} = u^a\nabla_a\Omega$. Therefore, even if $\Theta = 0$, the expansion scalar $\tilde{\Theta}$ in the conformally connected background does not vanish unless $u^a \nabla_a \Omega = 0$. As a result, the local conservation of the entropy current does not hold in $\tilde{g}_{ab}$, leading to the inequality $\tilde{\nabla}_a \tilde{S}^a > 0$, which indicates entropy production in the conformally transformed frame. Therefore, a non-viscous fluid in thermal equilibrium (\textit{i.e.} $q^a=0$) on $g_{ab}$ becomes a viscous fluid in thermal equilibrium (\textit{i.e.} $\tilde{q}^a=0$) on the conformally connected background $\Tilde{g}_{ab}$. This is reminiscent of the well-known fact that geodesics on the background $g_{ab}$ are no longer geodesics for the conformally connected background $\tilde{g}_{ab}$. This is because of the following reason. Looking at Eq. (\ref{BR5}), with the assumption that the geodesics in the $g_{ab}$ are affinely parametrized, providing $\dot{u}^a=0$. Thus use of Eq. (\ref{BR5}) yields
\begin{eqnarray}
  \dot{\Tilde{u}}_a &=& \Delta_{ab}\nabla^b\ln{\Omega} = \Tilde{u}_a\Tilde{u}_b\Tilde{\nabla}^b\ln{\Omega} + \Tilde{g}_{ab}\Tilde{\nabla}^b\ln{\Omega}~.
\end{eqnarray}   
Now, the first term proportional to $\Tilde{u}_a$ can be absorbed by doing an affine parametrization, as it arises when a geodesic is non-affinely parametrized. However, the second term spoils the geodesic character in the $\Tilde{g}_{ab}$ background. Therefore, the above one is no longer a geodesic. This particular term can be interpreted as a forcing term in the conformally connected background.

\section{Geometrothermodynamics}

In the previous sections, the relations among the equilibrium thermodynamic fluid parameters are obtained by analyzing the dynamics of the fluid on conformally connected backgrounds. However, since it is known that equilibrium thermodynamics can also be formulated through a geometric framework, it would be interesting to investigate how these geometric descriptions relate when defined on two conformally connected backgrounds, given the relations among the equilibrium thermodynamic variables. This equilibrium geometric description was developed by Quevedo through the use of Legendre-invariant metrics, which relate the equilibrium thermodynamic properties of a system using the formalism of differential geometry \cite{Quevedo_2007,quevedo2011fundamentalsgeometrothermodynamics}. Here we present a short summary for the sake of completeness before coming to our main discussion.

To relate the concepts of thermodynamics to the ideas of differential geometry, a $(2n+1)-$ dimensional thermodynamic phase space $\mathcal{T}$ is introduced with coordinates  $Z^A= {\left(\Phi,E^a,I^a\right)}$, where $\Phi$ is the thermodynamic potential, and $E^a$ and $I^a$ are the extensive and intensive variables respectively with $A=0,1,...,2n$ such that $\Phi = Z^0$, $E^a = Z^a$ and $I^a = Z^{n+a}$ for $a=1,2,....,n$.
Thereafter, a contact structure is defined over $\mathcal{T}$, given by 
\begin{equation}
    \Theta_G = d\Phi - \delta_{ab}I^a\,dE^b, \quad \delta_{ab} = \text{diag}(1,1,...,1).
    \label{G0}
\end{equation}
The pair $(\mathcal{T},\Theta_G)$ is called a contact manifold if $\mathcal{T}$ is differential and $\Theta_G$ satisfies $\Theta_G \wedge (d\Theta_G)^n \neq 0$. Further, the space of thermodynamic equilibrium states $\mathcal{E}$ is defined by the smooth mapping $\varphi:\mathcal{E}\longrightarrow \mathcal{T}$ given by 
\begin{equation}
    \varphi : (E^a) \longrightarrow (\Phi, E^a, I^a),
    \label{GG1}
\end{equation}
with $\Phi = \Phi(E^a)$.
The pullback $\varphi^{*}$ gives the condition 
\begin{equation}
    \varphi^{*}(\Theta_G) = \varphi^{*}( d\Phi - \delta_{ab} I^a \, dE^b) = 0.
    \label{GG2}
\end{equation}
The above condition on $\mathcal{E}$, leads to the first law of thermodynamics
\begin{equation}
d\Phi - \delta_{ab} I^a \, dE^b = 0.
\label{GG3}
\end{equation}

Furthermore, to describe a thermodynamic system, a \textit{thermodynamic metric} $G$ is introduced on $\mathcal{T}$, which satisfies two conditions. First, it is invariant under Legendre transformations, and secondly, ${G}$ induces an invariant metric ${g}$ on the equilibrium space $\mathcal{E}$ by the mapping
\begin{equation}
    \varphi^*(G) = g.
    \label{GG6}
\end{equation}
The invariance of the \textit{thermogeometric metric} $G$ is essential, as the description of a thermodynamic system must not depend upon the thermodynamic potential in consideration. The Legendre transformation is given by
\begin{equation}
\begin{aligned}
\{ Z^A \} &\longrightarrow \{ Z'^A \} = \{ \Phi', E'^a, I'^a \}, \\
\Phi' &= \Phi - \delta_{i j} E'^i I'^j, \quad E^i = -I'^i, \quad I^j = E'^j,  \\
&  \quad E^k = E'^k, \quad I^k = I'^k.
\end{aligned}
\label{GG7}
\end{equation}
This ends our short introduction to the idea of GTD.

Now, Quevedo \textit{et al.} in their work \cite{Quevedo_2007,quevedo2011fundamentalsgeometrothermodynamics}, on analyzing the Legendre invariance conditions, found a Legendre invariant metric, which is given by 
\begin{equation}
\begin{aligned}
G = \Theta_G^2 + \Lambda \left( E_a I_a \right)^{2k+1} dE^a dI^a,\\
\quad E_a = \delta_{ab}E^b, \quad I_a = \delta_{ab}I^b.
\end{aligned}
\label{G2}
\end{equation}
To discuss our results, we work with a particular choice by setting $\Lambda=1$ and $k = 0$, which does not affect the generality of our results. With the above choice, the Legendre invariant metric on $\mathcal{T}$ is given by
\begin{equation}
\begin{aligned}
G = \Theta_G^2 + \left( E_a I_a \right) dE^a dI^a,
\quad E_a = \delta_{ab}E^b, \quad I_a = \delta_{ab}I^b.
\end{aligned}
\label{g2}
\end{equation}
One can easily check that the above metric is invariant under the Legendre transformations given by Eq. (\ref{GG7}), which makes it suitable to discuss a thermodynamic system. Coming to our discussion, we work with $Z^A= {\left(\Phi,E^a,I^a\right)} = {(U,S,V,N,T,-p,\mu)}$, where $\Phi = U$, $E^a = (S,V,N)$ are extensive variables and $I^a = (T,-p,\mu)$ are the intensive variables corresponding to the extensive variables. One important point to note is that the above discussion is formulated in terms of total quantities, such as total energy, entropy, and particle number, rather than their corresponding densities. These variables $U, S,$ and $N$ are the proper volume integrals of their corresponding densities—$\epsilon, s,$ and $n$, respectively—as described in the context of fluid dynamics in BDNK theory, which says, $U$, $S$, and $N$ represent the total internal energy, entropy, and baryon number respectively.
Therefore, use of Eq. (\ref{G0}) provides the contact form on $\mathcal{T}$ as
\begin{equation}
\Theta_G = \mathrm{d}U + p \mathrm{d}V - T \mathrm{d}S -\mu \mathrm{d}N.
\label{G1}
\end{equation}
Then the Legendre invariant metric given by Eq. (\ref{g2}) for our choice of variables reduces to 
\begin{eqnarray}
    G = \Theta_G^2 + \left(TS\,dT\,dS + pV\,dp\,dV + \mu N\,d\mu\,dN\right).
    \label{G3}
\end{eqnarray}

An another important point to note is that the extensive variables are $(S, V, N)$, and thus, when projecting onto the equilibrium space $\mathcal{E}$, the associated metric $g$, which is given by \E{GG6}, would naturally be a function of $(S, V, N)$. However, since our main analysis focuses on the temperature $T$ and chemical potential $\mu$, we prefer to work in a framework where these are the primary variables. To achieve this, we perform a total Legendre transformation given by Eq. (\ref{GG7}), which exchanges the roles of the extensive and intensive variables. This is as follows:
\begin{equation}
\begin{aligned}
U' &= U - TS + pV - \mu N, \\
S' &= T, \quad V' = -p, \quad N' = \mu, \\
T' &= -S, \quad -p' = -V, \quad \mu' = -N~,
\end{aligned}
\label{G4}
\end{equation}
and
\begin{multline}
{Z'}^A = ({U'}, {S'}, {V'}, {N'}, {T'}, -{p'}, {\mu'}) \\
 = (U - TS + pV - \mu N,T,-p,\mu,-S,V,-N).
\label{G5}
\end{multline}
In this framework, the independent variables are ${E'}^a = (S', V', N') = (T, -p, \mu)$, and the corresponding conjugate intensive variables are ${I'}^a = (T', -p', \mu') = (-S, -V, -N)$. As previously noted, the metric $G$ remains invariant under Legendre transformations, meaning that its form is preserved. Consequently, the transformed metric $G'$ retains the same structure and is again given by Eq. (\ref{G3}).

 To analyze the thermodynamic properties of equilibrium states, we now project the transformed metric $G'$ (obtained by performing the Legendre transformations \E{G4}), which lies on the thermodynamic phase space $\mathcal{T}$, onto the equilibrium sub-manifold $\mathcal{E}$ described by $g'$.
On projecting $G'$ onto the equilibrium space $\mathcal{E}$, we have 
\begin{equation}
g' = TS\,dT\,dS + pV\,dp\,dV + \mu N\,d\mu\,dN~.
\label{G6}
\end{equation}
As previously mentioned, the extensive variables after the Legendre transformations are now $T, p,$ and $\mu$.  Moreover, on the space of equilibrium states $\mathcal{E}$, we also have the Gibbs-Duhem relation, $Vdp = S dT + N d\mu$. Thus, the use of the Gibbs-Duhem relation in \E{G6}, provides 
\begin{equation}
{g}' = TS\,dT\,dS + p\,S\,dV\,dT + p\,N\,dV\,d\mu + \mu N\,d\mu\,dN~.
\label{GGG1}
\end{equation}
Similarly, the thermogeometric metric describing the fluid corresponding to $\Tilde{g}_{ab}$ is given by:
\begin{equation}
\Tilde{g}' = \Tilde{T}\Tilde{S}\,d\Tilde{T}\,d\Tilde{S} + \Tilde{p}\,\Tilde{S}\,d\Tilde{V}\,d\Tilde{T} + \Tilde{p}\,\Tilde{N}\,d\Tilde{V}\,d\Tilde{\mu} + \Tilde{\mu} \Tilde{N}\,d\Tilde{\mu}\,d\Tilde{N}~.
\label{GGG2}
\end{equation}

Now, as shown in Section II, if a fluid is in thermal equilibrium within two conformally connected spacetimes, $g_{ab}(x)$ and $\tilde{g}_{ab}(x)$, related through Eq.~(\ref{BR2}), then the corresponding fluid parameters in these backgrounds are related by Eq.~(\ref{BR6}), Eq.~(\ref{BR20}), Eq.~(\ref{BR15}), and Eq.~(\ref{BR17}).
Now, given these relations among the thermodynamic parameters, it can be shown that the thermogeometric metrics describing the fluid given by $g'$ and $\Tilde{g}'$ corresponding to $g_{ab}(x)$ and $\Tilde{g}_{ab}(x)$ respectively are also conformally connected as explicitly shown below.

In order to show that \E{GGG1} and \E{GGG2} are conformally connected, let us look at how the 
entropy $\Tilde{S}$ and $S$ are related. Before proceeding, we note that the spatial volume elements in $(3+1)$-spacetime dimensions are related by $d\Tilde{V} = \Omega^3\,dV$. Also, we have 
\begin{equation}
    \Tilde{S} = \int \Tilde{s}\,d\Tilde{V}~,
    \label{GGG3}
\end{equation}
and similarly for $S$ with untilde quantities.
Then, the use of \E{BR17}, along with $d\Tilde{V} = \Omega^3\,dV$, yields
\begin{equation}
    \Tilde{S} = \int \Tilde{s}\,d\Tilde{V} = \int s\, dV =  S.
    \label{GGG4}
\end{equation}
The same was observed earlier in an explicit example \cite{Bhattacharya:2018xlq} related to the thermodynamics of horizons corresponding to two conformally connected gravitational theories.
Additionally, an exactly similar calculation provides $\Tilde{N} = N$. Further, it is important to note that the thermodynamic parameters—such as temperature—are related by Eq. (\ref{BR6}). During a thermodynamic process in the background $g_{ab}$, if the temperature changes from $T_i$ to $T_f$, then in the conformally connected background $\Tilde{g}_{ab}$, the corresponding temperatures transform as $\Tilde{T}_i = T_i/\Omega$ and $\Tilde{T}_f = T_f/\Omega$. Therefore, the temperature variation in $\Tilde{g}_{ab}$ is given by $d\Tilde{T} = \Tilde{T}_f - \Tilde{T}_i = (T_f - T_i)/\Omega = dT/\Omega$. A similar relation follows for the chemical potential $\mu$. 
Therefore, use of the above argument for the thermodynamic parameters along with \E{BR6}, \E{BR20}, \E{BR17}, and \E{GGG4}, into \E{GGG2} yields
\begin{equation}
\Tilde{g}' = \Omega^{-2} g ~.
    \label{G17}
\end{equation}
This highlights the fact that even the thermogeometric metrics $g'$ and $\Tilde{g}'$, which describe the fluid on $g_{ab}$ and $\Tilde{g}_{ab}$, respectively, maintain a conformal relationship under those relations among the equilibrium thermodynamic parameters.
This is probably expected as the fluid variables are just scaled by the conformal factor when going from $g_{ab}$ to $\tilde{g}_{ab}$, and geometries corresponding to thermodynamic relations are constructed through a Legendre invariant way.
In this sense, such an observation could be interpreted as a consistency check for the relations among thermodynamic parameters. Moreover, \E{BR9} can be provided an interesting perspective by analyzing the ``causal" structure of \E{G17} as discussed below.

It is known that the usual light cones of a spacetime remains invariant under conformal transformation. Also in our case we just observed that the metrics $g'$ and $\tilde{g}'$ are conformally connected (see Eq. (\ref{G17})). Therefore knowing this, we will investigate whether (\ref{BR9}) can be interpreted as a cause of such conformal relation.
As mentioned above, working with $G'$ and $g'$ switches the extensive and intensive variables, which provides us with $S(T, -p, \mu)$, $V(T, -p, \mu)$ and $N = (T, -p, \mu)$. Therefore we have
\begin{equation}
\begin{aligned}
dS &= \left( \frac{\partial S}{\partial T} \right)_{p, \mu} dT + \left( \frac{\partial S}{\partial p} \right)_{T, \mu} dp + \left( \frac{\partial S}{\partial \mu} \right)_{T, p} d\mu~; \\
dV &= \left( \frac{\partial V}{\partial T} \right)_{p, \mu} dT + \left( \frac{\partial V}{\partial p} \right)_{T, \mu} dp + \left( \frac{\partial V}{\partial \mu} \right)_{T, p} d\mu~; \\
dN &= \left( \frac{\partial N}{\partial T} \right)_{p, \mu} dT + \left( \frac{\partial N}{\partial p} \right)_{T, \mu} dp + \left( \frac{\partial N}{\partial \mu} \right)_{T, p} d\mu~.
\end{aligned}
\label{G7}
\end{equation}
Then using the Eq. (\ref{G7}) into Eq. (\ref{GGG1}), along with the Gibbs-Duhem relation $V dp = S dT + N d\mu$,  \E{GGG1} reduces to
\begin{equation}
g' = \left[ A (dT)^2 + 2B dT d\mu + C (d\mu)^2 \right]~,
\label{G9}
\end{equation}
where the coefficients $A,B,$ and $C$ are given by the following expressions
\begin{equation}
\begin{aligned}
A = & TS \left[ \left( \frac{\partial S}{\partial T} \right)_{p,\mu}
+ \frac{S}{V} \left( \frac{\partial S}{\partial p} \right)_{T,\mu} \right] \\
&+ pV \left[ \frac{S}{V} \left( \frac{\partial V}{\partial T} \right)_{p,\mu}
+ \frac{S^2}{V^2} \left( \frac{\partial V}{\partial p} \right)_{T,\mu} \right]~,
\end{aligned}
\label{G10}
\end{equation}
\begin{equation}
\begin{aligned}
B &=\frac{1}{2} \Bigg\{ 
TS \left[ \left( \frac{\partial S}{\partial \mu} \right)_{T,p}
+ \left( \frac{\partial S}{\partial p} \right)_{T,\mu} \frac{N}{V} \right] \\
&\hspace{2cm}+ \mu N \left[ \left( \frac{\partial N}{\partial T} \right)_{p,\mu}
+ \left( \frac{\partial N}{\partial p} \right)_{T,\mu} \frac{S}{V} \right] +  \\
& pV \left[ \left( \frac{\partial V}{\partial T} \right)_{p,\mu} \frac{N}{V}
+ 2 \left( \frac{\partial V}{\partial p} \right)_{T,\mu} \frac{SN}{V^2}
+ \left( \frac{\partial V}{\partial \mu} \right)_{T,p} \frac{S}{V} \right]
\Bigg\}~,
\end{aligned}
\label{G11}
\end{equation}
and 
\begin{equation}
\label{G12}
\begin{aligned}
C =\; & \mu N \left[ \left( \frac{\partial N}{\partial \mu} \right)_{T,p}
+ \left( \frac{\partial N}{\partial p} \right)_{T,\mu} \frac{N}{V} \right] \\
&+ pV \left[ \left( \frac{\partial V}{\partial \mu} \right)_{T,p} \frac{N}{V}
+ \left( \frac{\partial V}{\partial p} \right)_{T,\mu} \frac{N^2}{V^2} \right]~.
\end{aligned}
\end{equation}
Correspondingly, in similar fashion, the thermogeometric metric $\Tilde{g}'$ describing the fluid on $\Tilde{g}_{ab}$ given by {\E{GGG2} reduces to
\begin{equation}
\Tilde{g}' = \left[ \tilde{A} (d\Tilde{T})^2 + 2\Tilde{B}\, d\Tilde{T} \,d\Tilde{\mu} + \tilde{C} (d\Tilde{\mu})^2 \right]~.
\label{G13}
\end{equation}
where $\Tilde{A}$, $\Tilde{B}$, and $\Tilde{C}$ are given Eqs.~(\ref{G10})--(\ref{G12}) with tilde on every variable. 
 Now, on use of Eq. (\ref{BR6}), Eq. (\ref{BR20}) into Eq. (\ref{G13}) along with Eq. (\ref{G17}) and \E{G9}, it is fairly easy to obtain the relation between the metric coefficients of $g'$ and $\Tilde{g'}$
\begin{eqnarray}
&&\Tilde{A} =A,\,\,\,\,\Tilde{B} = B,\,\,\,\,\Tilde{C} = C.
\label{G15}
\end{eqnarray} 

Now to find the light cone, one needs to fix $\Tilde{g}'=0$. Then Eq. (\ref{G13}) yields
\begin{equation}
\frac{d\Tilde{\mu}}{d\Tilde{T}}= \frac{-\Tilde{B} \pm \sqrt{\Tilde{B}^2 - \Tilde{A}\Tilde{C}}}{\Tilde{C}}~.
\label{G14}
\end{equation}
Similarly, the light cone for $g'$ is given by the above structure with tilde variables replaced by untilde ones.
Then the use of \E{G15} and \E{G17} yields,
\begin{equation}
 \frac{d\Tilde{\mu}}{d\Tilde{T}}= \frac{d \mu}{ d T}~.   
 \label{G18}
\end{equation}
This shows that the slope of the ``light rays'' is same at each pair of point (one point $(\mu,T)$ on $g'$ and the corresponding one $(\tilde{\mu},\tilde{T})$ on $\tilde{g}'$) for these two thermogeometric metrics. This is expected as the metrics are conformally connected. So for a given $(\mu,T)$ on $g_{ab}(x)$ and corresponding $(\tilde{\mu}, \tilde{T})$ on $\tilde{g}_{ab}(x)$ the above relation implies the result given in $(\ref{BR9})$ at each spacetime event $x$. Hence, in summary, the conformal connection between spacetime metrics implies a conformal relation between thermogeometric descriptions of the fluid thermodynamic parameters, which effectively leads to (\ref{BR9}) and vice versa. 

\section{Conclusions and discussions}
The generalization of the TE relation and Klein's law to an arbitrary background is quite difficult. However, we have built up relations between thermodynamic variables of fluids, described on two conformally connected spacetimes. In particular, we found that if the fluid is in thermal equilibrium in both backgrounds, then the temperature and chemical potential for the conformally rescaled spacetime scale with the inverse of the conformal factor to those on the seed metric. Surprisingly, we observed that if Klein's law holds in the seed background—\textit{i.e.} if $\mu/T$ remains constant throughout the spacetime described by $g_{ab}$ (in that case $g_{ab}$ is static or stationary) then, as stated by Eq. (\ref{BR9}), it also holds in the rescaled background, even if the $\Tilde{g}_{ab}$ is neither static nor stationary. We further obtained the relations between other thermodynamic variables by demanding the conservation of baryon current in both backgrounds. Further, by using the idea of GTD, we also discussed the consistency of our results. We found that the Legendre invariant metrics describing the fluid on the space of equilibrium states $\mathcal{E}$ are themselves conformally connected. Moreover, we also provided an interesting way to analyze \E{BR9}, in terms of the ``causal" structure of these Legendre invariant metrics on $\mathcal{E}$. 

A similar direction was also invoked recently in \cite{Faraoni:2023gqg}, however, the present analysis bears differences. It is worthy to mention the noticeable features of the results obtained so far. 
\begin{itemize}
    \item The earlier analysis is restricted in the sense that the seed metric $g_{ab}$ is static. However, our analysis is valid for any arbitrary conformally connected backgrounds that are solutions to Einstein's equations.
    \item The previous derivation for $\Tilde{T}=T/\Omega$ assumes a particular choice for $u^a$ given by $u_a=-N\nabla_a t$. We did not invoke any such choice of fluid four-velocity; as of now, it is completely arbitrary.
    \item The relation between the thermodynamic variables was discussed in the context of an ideal fluid. We found that the same can be obtained for dissipative fluids by demanding the baryon current conservation and the absence of heat fluxes in both backgrounds.
    \item Further, we were able to extend the analysis by providing the relation between the chemical potentials in both backgrounds. This was not addressed in the literature.
    \item Interestingly, we find that Klein's law remains valid in the conformally connected background if it holds in the static or stationary seed background, even when the conformally connected background is neither static nor stationary.
    \item Contrary to earlier attempts, all the results have been developed through a much robust first order fluid description and therefore the conclusions are more reliable.
    \item We elaborate all these results through GTD, which was absent in the literature. 
\end{itemize}
Therefore, the present analysis illuminates the generality of the scaled relations among the thermodynamic quantities on two conformally connected spacetimes. Also, it complements Dicke's heuristic argument.

However, our results have certain limitations. As the stability and causality of BDNK theory have only been checked for Einstein's theory (see \cite{PhysRevX.12.021044}), our results are valid only for backgrounds that are solutions to Einstein's equation. Since with other theories of gravity the stability and causality of the fluid description have to be checked separately, the present results may not be trusted outside Einstein's theory. One scenario has been discussed in the context of scalar-tensor theory which has an extra scalar degree of freedom $\phi$ in \cite{Karolinski:2024ukr}. The authors found exactly the same relation for temperature. In their analysis, they have assumed the scalar field $\phi$ to be time-independent, which makes the conformal factor time-independent. Thus, both the seed background and the conformally connected background are static. It would be fascinating to generalize these ideas to arbitrary backgrounds.

\begin{acknowledgments}
The work of BR is supported by the University Grants Commission (UGC), Government of India, under the scheme Junior Research Fellowship (JRF). BRM is supported by Science and Engineering Research Board (SERB), Department of Science $\&$ Technology (DST), Government of India, under the scheme Core Research Grant (File no. CRG/2020/000616).
\end{acknowledgments}

\appendix
\section*{Appendix}
\section{Steps leading to Eq. (\ref{BR6}) and Eq. (\ref{BR20})}\label{App1}
In this section, we provide the detailed calculations that lead to our main results. In the BDNK theory, the expressions for the heat currents are given by Eq. (\ref{D4}) and Eq. (\ref{D5}) in the seed metric $g_{ab}$ and its conformally connected background $\Tilde{g}_{ab}$, respectively. 
Let us begin with the general relationships between temperatures and chemical potentials in the conformally connected backgrounds:
\begin{equation}
\Tilde{T}=\frac{T}{\Omega^{z_2}};\,\,\,\,\,\,\,\,\,\,\,\,\,\,\,\,\, \Tilde{\mu}=\frac{\mu}{\Omega^{z_3}}~,
\label{E1}
\end{equation}
where $z_2$ and $z_3$ are some real numbers. These are to be determined. The above choices are consistent with the fact that for $\Omega=1$, we must have $\Tilde{T}  = T$ and $\tilde{\mu} = \mu$. Note that for this condition, one can also add terms which are derivatives of $\Omega$; however, those must have the dimension of temperature for the first relation and chemical potential for the other one. However, we prefer to choose the above simple relations and investigate if they lead to any meaningful objective.
Using Eq. (\ref{E1}) into Eq. (\ref{D5}) along with $\Tilde{\Delta}^{ab} = \Omega^{-2}\Delta^{ab}$ and Eq. (\ref{BR5}), yields
\begin{multline}
{\Tilde{q}}^a = \Omega^{-2}\Big[\Tilde{\alpha}\,\Omega^{z_2-z_3}\Delta^{ab}\nabla_b\left(\frac{\mu}{T}\right) + \Tilde{\beta}\Delta^{ab}\left(\nabla_b\ln{T} + \dot{u}_b\right)\\
+ \Tilde{\beta}\Delta^{ab}\nabla_b\ln{(\Omega)^{1-z_2}} + \Tilde{\alpha}\,\frac{\mu}{T}\Delta^{ab}\nabla_b(\Omega)^{z_2-z_3}\Big]~.
\label{E4}
\end{multline}
Further use of Eq. (\ref{imp}) provides
\begin{multline}
\Omega^{z_1} q^a =  \Omega^{-2}\Big[\Tilde{\alpha}\,\Omega^{z_2-z_3}\Delta^{ab}\nabla_b\left(\frac{\mu}{T}\right) + \Tilde{\beta}\Delta^{ab}\left(\nabla_b\ln{T} + \dot{u}_b\right)\\
+ \Tilde{\beta}\Delta^{ab}\nabla_b\ln{(\Omega)^{1-z_2}} + \Tilde{\alpha}\,\frac{\mu}{T}\Delta^{ab}\nabla_b(\Omega)^{z_2-z_3}\Big]~.
\label{E5}
\end{multline}
Substituting the expression for $q^a$, given by Eq. (\ref{D4}), in the left hand side of (\ref{E5}) we have
\begin{multline}
\Omega^{z_1 + 2} \Big[\alpha\Delta^{ab} \nabla_b\left(\frac{\mu}{T}\right)+\beta \Delta^{ab}\left(\nabla_b \ln{T}+\dot{u}_b\right)\Big]\\
= \Big[\Tilde{\alpha}\,\Omega^{z_2-z_3}\Delta^{ab}\nabla_b\left(\frac{\mu}{T}\right) + \Tilde{\beta}\Delta^{ab}\left(\nabla_b\ln{T} + \dot{u}_b\right)\Big]\\
+ \Big[\Tilde{\beta}\Delta^{ab}(1-z_2)\nabla_b\ln{(\Omega)} + \Tilde{\alpha}\,\frac{\mu}{T}\Delta^{ab}(z_2 - z_3)(\Omega)^{z_2-z_3 -1}\nabla_b\Omega\Big]~.
\label{E6}
\end{multline}

Now, since $T$ and $\mu$ are independent variables, making $T$ and $\mu/T$ also independent (see \cite{Statement}), Eq. (\ref{E6}) is valid for an arbitrary conformal factor provided the following relations are satisfied: 
\begin{eqnarray}
&&\alpha\,\Omega^{z_1+ 2} = \Tilde{\alpha}\,\Omega^{z_2 - z_3}~;
\label{E7}
\\
&&\beta \, \Omega^{z_1 +2} = \Tilde{\beta}~; 
\label{E8}
\end{eqnarray}
and
\begin{equation}
\Tilde{\beta}\Delta^{ab}(1-z_2)\nabla_b\ln{(\Omega)} + \Tilde{\alpha}\,\frac{\mu}{T}\Delta^{ab}(z_2 - z_3)(\Omega)^{z_2-z_3 -1}\nabla_b\Omega = 0~.
\label{E10}
\end{equation}
The last one is re-expressed as
\begin{equation}
 \left(\Tilde{\beta}(1-z_2) + \Tilde{\alpha}\,\frac{\mu}{T}(z_2 - z_3)\Omega^{z_2-z_3}\right)\Delta^{ab}\nabla_b\Omega = 0~.
\label{E12}
\end{equation}
For the satisfaction of (\ref{E12}), we can have either
\begin{equation}
\left(\Tilde{\beta}(1-z_2) + \Tilde{\alpha}\,\frac{\mu}{T}(z_2 - z_3)\Omega^{z_2-z_3}\right)=0~,
\label{E12n}
\end{equation}
or $\Delta^{ab}\nabla_b\Omega=0$. The latter one shows one must have $\nabla_b\Omega \propto u_b$, which imposes a restriction on the conformal factor as well as on the choice of $4$-velocity. Therefore, one loses the arbitrariness, and hence we discard this possibility.
Under these circumstances (\ref{E12n}) is the only choice which, after use of Eq. (\ref{E7}) and Eq. (\ref{E8}) reduces to
\begin{equation}
(1-z_2) +\Big[\frac{\alpha \mu}{\beta T}\Big](z_2-z_3)= 0~.
\label{E13}
\end{equation}
The term in the third bracket is a function of spacetime coordinates, whereas other terms are not. Since Eq. (\ref{E13}) holds at every spacetime point in thermal equilibrium, the above equation leads to 
\begin{equation}
z_2 =1, \,\,\,\  z_3=z_2~.
\label{E17}
\end{equation}
This demonstrates that the temperature and chemical potential in the two conformally connected backgrounds are related by Eq. (\ref{BR6}) and Eq. (\ref{BR20}).
Furthermore, from Eq. (\ref{E7}) and Eq. (\ref{E8}) along with the use of Eq. (\ref{E17}), we have Eq. (\ref{E18}). It may be noted that all the required relations were obtained without knowing the explicit value of $z_1$, and so $z_1$ remains a non-vanishing arbitrary real number.

\bibliographystyle{apsrev}

\bibliography{bibtexfile}

\begin{thebibliography}{47}
\bibitem[{Sta()}]{Statement}
\bibinfo{note}{For a fluid description, one needs five independent variables.
  BDNK theory is formalised by considering three components of velocity (as
  $u^a$ is constraint by $u^au_a = -1$), $\epsilon$ and $n$
  \cite{PhysRevX.12.021044}. However, using the relation $Tds = d\epsilon - \mu
  dn$ and the equation of state, one can show that $\epsilon(T,\mu)$ and
  $n(T,\mu)$. Therefore $\epsilon$ and $n$ can be expressed in terms of $T$ and
  $\mu$ and hence the fluid description can also be given using three velocity
  components, $T$ and $\mu$ as independent variables.}
\end{thebibliography}

\end{document}